\pdfoutput=1

\documentclass[aps,prl,10pt,twocolumn,showpacs,superscriptaddress,footnoteinbib]{revtex4}
\usepackage{amsmath}
\usepackage{latexsym}
\usepackage{amssymb}
\usepackage{graphics,epstopdf}

\begin{document}

\title{Witness of mixed separable states useful for entanglement creation}
\author{Nirman Ganguly}
\thanks{nirmanganguly@gmail.com}
\affiliation{Department of Mathematics, Heritage Institute of Technology, Kolkata-700107,  India}
\affiliation{S. N. Bose National Centre for Basic Sciences, Salt Lake,
Kolkata-700098, India}
\author{Jyotishman Chatterjee}
\thanks{jyotishman$\_$c@yahoo.co.in}
\affiliation{Department of Mathematics, Heritage Institute of Technology, Kolkata-700107,  India}
\author{A. S. Majumdar}
\thanks{archan@bose.res.in}
\affiliation{S. N. Bose National Centre for Basic Sciences, Salt Lake,
Kolkata-700098, India}

\date{\today}

\begin{abstract}
Absolutely separable states form a special subset of the set of all separable 
states, as they remain separable under any global unitary transformation unlike
other separable states. In this work we consider the set of absolutely 
separable bipartite states and show that it is convex and compact in any 
arbitrary dimensional Hilbert space. Through a generic approach of construction
of suitable hermitian operators we prove the completeness of the 
separation axiom for identifying any separable state that is not absolutely 
separable. We demonstrate the action of such witness operators in different 
qudit systems.  Examples of mixed separable systems are provided, pointing out 
the utility of the witness in entanglement creation using quantum gates. 
Decomposition of witnesses in terms of spin operators or photon
polarizations facilitates their measureability for qubit states. 
\end{abstract}


\maketitle

\paragraph{Introduction.\textemdash} 

The problem whether a quantum state is separable or 
entangled remains one of the most involved problems in quantum information 
science, which is underlined by the observation that the separability problem 
is NP hard \cite{gurvits1}. A pure quantum state $\vert \psi \rangle$ is   
separable if it can be written in the product form as 
$\vert \psi \rangle = \vert \psi_{1} \rangle \otimes \vert \psi_{2} \rangle$. A 
mixed quantum state $\varrho_{sep}$ is separable if it can be written 
as $\varrho_{sep} = \sum_{i=1}^{k}p_{i}\vert e_{i},f_{i} \rangle \langle e_{i},f_{i} \vert $, where $\vert e_{i},f_{i} \rangle$ are product states and 
$p_{i}\geq 0, \sum_{i=1}^{k}p_{i}=1$. States which are not separable are called 
entangled. In lower dimensions, specifically in $2 \otimes 2$ and $2 \otimes 3$
 a state is separable if and only if it has a positive partial 
transpose \cite{peres,horodecki}. However, in higher dimensions there exist 
entangled states with positive partial transpose \cite{bound}. 

The theory of 
entanglement witnesses \cite{horodecki,terhal,review} provides a useful 
procedure to check whether a state is entangled. Entanglement witnesses $W$ 
are hermitian operators with at least one negative eigenvalue and satisfy the 
inequalities (i) $Tr(W \varrho_{sep}) \geq 0, \forall$ separable states 
$\varrho_{sep}$ and (ii) $Tr(W\varrho_{ent}) < 0$ for at at least one entangled 
state $\varrho_{ent}$. Entanglement witnesses can be used to detect the 
presence of entanglement experimentally \cite{review,barbieri,wiec}. The strength of 
the theory of entanglement witnesses is also due to its completeness, which 
asserts that if a state is entangled there will always be a witness that 
detects it \cite{horodecki}.

Entanglement witnesses (EW) constitute an application of a more general 
framework which comes under the domain of the geometric form of the celebrated 
Hahn-Banach theorem in functional analysis \cite{holmes}. The theorem states 
that if a set is convex and compact, then any point lying outside the set can 
be separated by a hyperplane.  The 
separation axiom has been also utilized in the inception of teleportation 
witnesses \cite{telwit}, which identify useful entangled states for quantum 
teleportation. Further, analogous to entanglement witnesses, work in the 
direction of constructing optimal 
\cite{telwit2} and complete \cite{telwitcom} 
teleportation witnesses has also been performed.  

An intriguing feature of the set of separable states is concerning the problem
of separability from spectrum \cite{knill}. This problem calls for a 
characterization of those separable states $\sigma$ for which 
$U\sigma U^{\dagger}$ is also separable for all unitary operators $U$. A possible
approach towards this end is to find constraints  on the eigenvalues of 
$\sigma$ such that it remains separable under any factorization of the 
corresponding Hilbert space.  The states that are separable from spectrum are 
also termed as absolutely separable states \cite{kus}, i.e., a separable state 
$\sigma$ is called absolutely separable if $U \sigma U^{\dagger}$ remains 
separable for any unitary operator $U$. There exists a ball of known radius 
centered at the maximally mixed state $\frac{1}{mn}(I \otimes I)$ 
(for $mn \times mn$ density matrices), where all the states within the ball 
are absolutely separable \cite{openball}. However, there exist absolutely 
separable states outside this ball too\cite{ishizaka}.

The problem of separability from spectrum was first handled  in the case of 
$2 \otimes 2$ systems \cite{verstraete}, where it was shown that $\sigma$ is 
absolutely separable if and only if (iff) its eigenvalues satisfy 
$\lambda_{1} \leq \lambda_{3} + 2\sqrt{\lambda_{2}\lambda_{4}}$, the eigenvalues 
being in the descending order. One closely related problem is the 
characterization of the states which have positive partial transpose (PPT) 
from spectrum, i.e., the states $\sigma_{ppt}$ with the property that 
$U \sigma_{ppt} U^{\dagger}$ is PPT for any unitary operator $U$. It was shown 
in \cite{hildebrand} that $\sigma_{ppt} \in D(H_{2} \otimes H_{n})$($D(X)$ represents the bounded linear operators acting on $X$) is PPT from 
spectrum iff its eigenvalues obey $\lambda_{1} \leq \lambda_{2n-1} +2\sqrt{\lambda_{2n-2}\lambda_{2n}}$. A recent progress in this problem was reported in 
\cite{johnston}, where it was shown that separability from spectrum is 
equivalent to PPT from spectrum for states living in $D(H_{2} \otimes H_{n})$. 

The generation of entanglement from separable states is
one of the leading experimental frontiers at present \cite{generation}. As 
absolutely separable states remain separable under global unitary operations, 
such states cannot be used as the initial input states for entanglement
generation. Though pure product states are not absolutely separable,
 the same is not true for mixed separable states which become absolutely 
separable after crossing a given amount of mixedness \cite{thirring}. Given
the ubiquity of environmental interactions in turning pure states into mixed
ones, it is of practical importance to determine whether a possessed state 
is eligible to be used as input for entanglement generation. The utility of 
mixed separable states which are not absolutely separable was highlighted 
in \cite{ishizaka} for the generation of maximally entangled mixed states.  
Mixed separable states from which entanglement can be created have also been
studied in other works \cite{mixsep}. 

Quantum gates have been employed to generate entanglement, especially in the 
context of quantum computation where unitary gates operate on qubits to
perform information processing.  Much work has been devoted to study the 
entangling capacity of unitary gates \cite{hybrid,entangling}.
Quantum algorithms use pure product states which can be turned into maximally 
entangled states using global unitary operations. However, if the state is
maximally mixed no benefit can be drawn from it as they remain invariant under 
global unitary operations. States in some vicinity of the maximally mixed 
state also remain  separable as noted in \cite{openball}. On the other hand,
separable mixed states may have possible implications in nuclear magnetic 
resonance quantum computation \cite{nmr}. However, states not close to the 
maximally mixed state may be useful for entanglement creation. So, it is 
important to study what happens when one moves from one extreme of a maximally 
mixed state to the other, i.e., a pure product state within the set of all 
separable states.

Given the immense significance of mixed separable states as stated above, 
we present here systematic proposal to identify separable states which are not 
absolutely separable. Our objective in this work is somewhat different from
the approach seeking to impose restrictions on the spectrum of absolutely
separable states \cite{verstraete,hildebrand,johnston}. Our motivation here
is to identify those separable states which are not absolutely separable, i.e.,
 the separable states $\chi$ for which $U \chi U^{\dagger}$ is 
entangled for some unitary operator $U$. To this end, we characterize the set 
of all absolutely separable states in any finite dimensional bipartite system 
as convex and compact. This enables one to  construct hermitian operators 
which identify separable states that are not absolutely separable in any 
arbitrary dimension Hilbert space. We propose a universal method of 
construction of such a witness operator and illustrate its action on states in 
different dimensions with a prescription for general two-qudit systems. 
Examples of unitary operations presented here include entangling gates
such as the celebrated {\it CNOT} (controlled {\it NOT}) gate, clearly 
distinguishing between absolutely separable states that remain separable
under global unitary operations from other states which get entangled. 
We show that the witness operators can be decomposed in terms of spin
operators and locally measureable photon polarizations for qubit states,
in order to facilitate their experimental realization.

\paragraph{Existence, construction and completeness of witness.\textemdash}

We begin with some notations and definitions
needed for our analysis. The density matrices that 
we consider here belong to any arbitrary dimension bipartite system, i.e., 
$\rho \in D(H_{m} \otimes H_{n})$. We denote by $\mathbf{S}$, the set of all 
separable states, i.e., $\mathbf{S}=\lbrace \rho : \rho $ is separable$\rbrace$,
 and the set of all absolutely separable states by $\mathbf{AS} = \lbrace \sigma \in \mathbf{S}: U \sigma U^{\dagger} $ is separable $\forall $ unitary operators
 $U \rbrace $. One can easily see that $\mathbf{AS}$ forms a non-empty subset of $\mathbf{S}$, as $\frac{1}{mn}(I \otimes I) \in \mathbf{AS}$. A point $x$ is 
called a limit point of a set $A$ if each open ball 
centered on $x$ contains at least
 one point of $A$ different from $x$. The set is closed if it contains each of 
its limit points \cite{simmons}. \\
\textbf{Theorem:} $\mathbf{AS}$ is a convex and compact subset of $\mathbf{S}$\\
\textbf{Proof:} \textit{$\mathbf{AS}$ \textbf{is convex}:} Let $\sigma_{1},\sigma_{2} \in \mathbf{AS}$ and $\sigma=\lambda \sigma_{1}+(1-\lambda)\sigma_{2}$, where $\lambda \in [0,1]$. Consider now, an arbitrary unitary operator $U$. 
Therefore,
\begin{equation}
U\sigma U^{\dagger}=\lambda U \sigma_{1} U^{\dagger} + ( 1- \lambda)  U \sigma_{2} U^{\dagger} = \lambda \sigma_{1}^{'} + ( 1- \lambda) \sigma_{2}^{'}
\label{convex1}
\end{equation}
where $\sigma_{i}^{'}=U \sigma_{i} U^{\dagger},i=1,2$. $\sigma_{1}^{'},\sigma_{2}^{'} \in \mathbf{S}$ as $\sigma_{1},\sigma_{2} \in \mathbf{AS}$ and now since $\mathbf{S}$ is convex , $U\sigma U^{\dagger} \in \mathbf{S}$ which implies $\sigma \in \mathbf{AS}$. Hence $\mathbf{AS}$ is convex.\\
\indent \textit{$\mathbf{AS}$ \textbf{is compact}:} Consider an arbitrary limit point $\theta$ of $\mathbf{AS}$ ($\mathbf{AS}$ will always have a limit point, 
for example, in the neighbourhood of the identity there are other absolutely 
separable states). The same $\theta$ must also be a limit point of $\mathbf{S}$ as $\mathbf{AS} \subset \mathbf{S}$ . Thus $\theta \in \mathbf{S}$, 
because $\mathbf{S}$ is closed. Now, let us inductively construct a 
sequence $\lbrace \theta_{n} \rbrace$ of distinct states from $\mathbf{AS}$ such that $\theta_{n} \rightarrow \theta$ as follows:
\begin{eqnarray}
 \theta_{1} & \in & B_{1}(\theta) \cap \mathbf{AS}, \>\>
\theta_{1}\neq \theta, \nonumber\\ \theta_{2} & \in & B_{\frac{1}{2}}(\theta) \cap \mathbf{AS}, 
\>\> \theta_{2}\neq \theta,\theta_{1} \nonumber\\
... & \in & .............. \nonumber\\
... & \in & .............. \nonumber\\
\theta_{n} & \in & B_{\frac{1}{n}}(\theta) \cap \mathbf{AS}, \>\> \theta_{n}\neq \theta,\theta_{1},\theta_{2}...\theta_{n-1}
\label{sequence}
\end{eqnarray}
 Here $B_{r}(\theta)$ denotes an open ball of radius $r$ centered at $\theta$. 
(This construction is possible because each neighbourhood of $\theta$ contains 
infinitely many points of $\mathbf{AS}$, $\theta$ being a limit point of $\mathbf{AS}$). For the above mentioned choice of $\theta_{n}$'s, evidently $\theta_{n} \rightarrow \theta$. Now, if we choose any unitary operator $U$, 
then $U \theta_{n} U^{\dagger} \rightarrow U \theta U^{\dagger}$.
Again, $U \theta_{n} U^{\dagger} \in \mathbf{S}$ for each $n \geq 1$, as $\theta_{n} \in \mathbf{AS}$. Since $\mathbf{S}$ is a closed set, it must contain the 
limit of the sequence $\lbrace U\theta_{n}U^{\dagger} \rbrace$, which 
is $U\theta U^{\dagger}$. Hence, $U\theta U^{\dagger} \in \mathbf{S}$, for 
arbitrary choice of the unitary operator $U$. Therefore, $\theta \in \mathbf{AS}$ as we already have $\theta \in \mathbf{S}$. Since $\theta$ is an arbitrary 
limit point of $\mathbf{AS}$, one can conclude that $\mathbf{AS}$ contains all 
its limit points, thereby implying that $\mathbf{AS}$ is closed \cite{simmons}.
 As any closed subset of a compact set is compact \cite{simmons}, one concludes
that $\mathbf{AS}$ is compact because $\mathbf{S}$ is compact. Hence, the 
theorem. $\square$

In view of the theorem above, we now formally define a hermitian operator $T$ 
which identifies separable but not absolutely separable states through the following two inequalities:
\begin{equation}
Tr(T\sigma) \geq 0, \>\> \forall \sigma \in \mathbf{AS}
\label{inequality1}
\end{equation}
\vskip -0.80cm
\begin{equation}
\exists \chi \in \mathbf{S}-\mathbf{AS}, ~~ \mathrm{s.t.} ~~ Tr(T\chi) < 0
\label{inequality2}
\end{equation}
Therefore, $T$ identifies those separable states $\chi$ that become 
entangled under 
some global unitary operation.

Consider $\chi \in \mathbf{S}-\mathbf{AS}$. There exists a unitary operator 
$U_{e}$ such that $U_{e}\chi U_{e}^{\dagger}$ is entangled. Consider an 
entanglement witness $W$ that detects $U_{e}\chi U_{e}^{\dagger}$, i.e., 
$Tr(W U_{e}\chi U_{e}^{\dagger}) < 0$. Using the cyclic property of the trace, 
one  obtains $Tr(U_{e}^{\dagger}W U_{e}\chi ) < 0$. We thus claim that 
\begin{equation}
T=U_{e}^{\dagger}W U_{e}
\label{witop}
\end{equation}
 is our desired operator. To see that it satisfies inequality (\ref{inequality1}), we consider its action on an arbitrary absolutely separable state $\sigma$. 
We have $Tr(T\sigma)=Tr(U_{e}^{\dagger}W U_{e} \sigma)=Tr(W U_{e} \sigma U_{e}^{\dagger} )$. As $\sigma $ is absolutely separable, $U_{e} \sigma U_{e}^{\dagger}$ is a 
separable quantum state, and since $W$ is an entanglement witness, $ Tr(W U_{e} \sigma U_{e}^{\dagger} ) \geq 0$. This implies that $T$ has a non-negative 
expectation value on all absolutely separable states $\sigma$. The completeness
of the separation axiom follows from the completeness of entanglement witness,
{\it viz.}, for any entangled state $U_{e}\chi U_{e}^{\dagger}$, there always
exists a witness $W$ \cite{horodecki}.
Thus, if $\chi$ is a separable but not absolutely separable state, then one 
can always construct an operator $T$ in the above mentioned procedure which 
distinguishes $\chi$ from absolutely separable states.

\paragraph{Illustrations.\textemdash}

As the first example consider the separable state in $D(H_{2} \otimes H_{2})$ 
given by \cite{thirring}
\begin{equation}
\chi_{2 \otimes 2} = \frac{1}{4}\left(%
\begin{array}{cccc}
  1 & 0 & 0 & 1 \\
  0 & 1 & 1 & 0  \\
  0 & 1 & 1 & 0  \\
  1 & 0 & 0 & 1  \\
\end{array}%
\right)
\end{equation}
 which becomes entangled on application of the unitary operator 
\begin{equation}
U_{1}=\frac{1}{\sqrt{2}}\left(%
\begin{array}{cccc}
  1 & 0 & 0 & 1 \\
  0 & \sqrt{2} & 0 & 0  \\
  0 & 0 & \sqrt{2} & 0  \\
  -1 & 0 & 0 & 1  \\
\end{array}%
\right)
\end{equation}
The entanglement witness 
\begin{equation}
W_{1}=\left(%
\begin{array}{cccc}
  c^{2} & 0 & 0 & 0 \\
  0 & 0 & -c & 0  \\
  0 & -c & 0 & 0  \\
  0 & 0 & 0 & 1  \\
\end{array}%
\right)
\end{equation}
with  $c=\frac{1}{\sqrt{2}+1}$ detects the entangled 
state $U_{1}\chi_{2 \otimes 2}U_{1}^{\dagger}$. Hence, the operator 
\begin{equation}
T_{1}=U_{1}^{\dagger}W_{1} U_{1}
\end{equation}
 gives $Tr(T_{1}\chi_{2 \otimes 2}) < 0$, detecting 
$\chi_{2 \otimes 2}$ to be a state which is not absolutely separable.

Next, consider the following separable density matrix $\chi_{2 \otimes 4} \in D(H_{2} \otimes H_{4})$:
\begin{equation}
\chi_{2 \otimes 4} = \left(%
\begin{array}{cccccccc}
  1/4 & 0 & 1/4 & 0 & 0 & 0 & 0 & 0\\
  0 & 0 & 0 & 0 & 0 & 0 & 0 & 0\\
  1/4 & 0 & 1/4 & 0  & 0 & 0 & 0 & 0\\
  0 & 0 & 0 & 0 & 0 & 0 & 0 & 0\\
  0 & 0 & 0 & 0 & 0 & 0 & 0 & 0\\
  0 & 0 & 0 & 0 & 0 & 1/4 & 0 & 1/4\\
  0 & 0 & 0 & 0 & 0 & 0 & 0 & 0\\
  0 & 0 & 0 & 0 & 0 & 1/4 & 0 & 1/4\\
\end{array}%
\right)
\end{equation}
The state $\chi_{2 \otimes 4}^{e}=U_{2} \chi_{2 \otimes 4} U_{2}^{\dagger}$, is 
entangled due to the unitary operator 
\begin{equation}
U_{2} = \frac{1}{\sqrt{2}}\left(%
\begin{array}{cccccccc}
  1 & 0 & 0 & 0 & 0 & 0 & 0 & 1\\
  0 & \sqrt{2} & 0 & 0 & 0 & 0 & 0 & 0\\
  0 & 0 & \sqrt{2} & 0  & 0 & 0 & 0 & 0\\
  0 & 0 & 0 & \sqrt{2} & 0 & 0 & 0 & 0\\
  0 & 0 & 0 & 0 & \sqrt{2} & 0 & 0 & 0\\
  0 & 0 & 0 & 0 & 0 & \sqrt{2} & 0 & 0\\
  0 & 0 & 0 & 0 & 0 & 0 & \sqrt{2} & 0\\
  -1 & 0 & 0 & 0 & 0 & 0 & 0 & 1\\
\end{array}%
\right)
\end{equation}
Therefore, the operator
$T_{2}=U_{2}^{\dagger} W_{2} U_{2}$
detects the state $\chi_{2 \otimes 4}$ 
as a separable but not absolutely separable state, where $W_{2}$ is the 
entanglement witness for the entangled state $\chi_{2 \otimes 4}^{e}$, given by
$ W_{2}=Q^{T_{B}}$with $Q$ being a projector on $\vert 10 \rangle - \vert 01 \rangle$.

It is hard to classify states separable from spectrum in dimensions other 
than $2 \otimes n$, due to the absence of suitable methodology in the existing
literature. However, through our approach of witnesses we can identify states 
which are not absolutely separable in any arbitrary dimension. This is 
demonstrated through the following illustration using a density matrix 
$\in D(H_{3} \otimes H_{3})$. Let us consider the isotropic state 
\begin{equation}
\chi_{3 \otimes 3}=\alpha\vert \phi_{3}^{+} \rangle \langle \phi_{3}^{+} \vert + \frac{1-\alpha}{9}I,
\end{equation}
 where $\vert \phi_{3}^{+} \rangle = \frac{1}{\sqrt{3}}(\vert 00 \rangle + \vert 11 \rangle + \vert 22 \rangle)$. This state is separable 
for $-\frac{1}{8}\leq \alpha \leq \frac{1}{4}$ \cite{isotropic}. It is observed 
that the unitary operator $U_{3}=I-(\frac{\sqrt{2}-1}{\sqrt{2}})(\vert 00 \rangle \langle 00 \vert + \vert 22 \rangle \langle 22 \vert) + \frac{1}{\sqrt{2}}(\vert 00 \rangle \langle 22 \vert - \vert 22 \rangle \langle 00 \vert) $ converts 
$\chi_{3 \otimes 3}$ to an entangled state $\chi_{3 \otimes 3}^{e}$ in the range 
$\alpha \in (\frac{1}{1+3\sqrt{2}},\frac{1}{4}]$. So again, the operator 
$T_3 = U_3^{\dagger}W_3U_3$
 detects $\chi_{3 \otimes 3}$ as a state that is not absolutely separable. Here 
$W_{3}$ is the entanglement witness that detects $\chi_{3 \otimes 3}^{e}$, given by
$ W_{3}=(\vert \eta \rangle \langle \eta \vert)^{T_{B}}$
with $\vert \eta \rangle$ being the eigenvector of $(\chi_{3 \otimes 3}^{e})^{T_{B}}$
corresponding to the eigenvalue $-\frac{1}{9}\alpha + \frac{1}{9}-\frac{\sqrt{2}}{3}\alpha$.

Let us now present a construction of the witness operator for general qudit
states. The form of the operator in $d \otimes d$ dimensions is obtained by
considering the following unitary operator:
\begin{eqnarray}
U_{d \otimes d}=I-(\frac{\sqrt{2}-1}{\sqrt{2}})A + \frac{1}{\sqrt{2}}B
\end{eqnarray}
where $A=\vert 00 \rangle \langle 00 \vert + \vert d-1,d-1 \rangle \langle d-1,d-1\vert$ and $B=\vert 00 \rangle \langle d-1,d-1 \vert - \vert d-1,d-1 \rangle \langle 00 \vert$. Consider now the mixed separable state 
\begin{eqnarray}
\chi_{d \otimes d}=\frac{1}{4}\vert 00 \rangle \langle 00 \vert + \frac{3}{4}\vert d-1,d-1 \rangle \langle d-1,d-1 \vert
\end{eqnarray}
 The state $U_{d \otimes d}\chi_{d \otimes d}U_{d \otimes d}^{\dagger}$ is entangled as 
detected by the witness 
$W_{d \otimes d}=\frac{1}{d}I-\vert P \rangle \langle P \vert$,
where $P$ is the projector on the maximally entangled state 
$\frac{1}{\sqrt{d}}\sum_{i=0}^{d-1} \vert ii \rangle$. Therefore, in 
$d \otimes d$ dimensions the operator 
$T_{d \otimes d}=U_{d \otimes d}^{\dagger} W_{d \otimes d} U_{d \otimes d}$
 detects $\chi_{d \otimes d}$ as a state which is not absolutely separable.

\paragraph{Entanglement creation using quantum gates.\textemdash}

Let us now consider some examples of unitary quantum gates which can produce
entanglement by acting on bipartite separable states.
Since the construction presented above is valid for any arbitrary dimension, 
let us consider a case in $d_{1} \otimes d_{2}$ dimensions where 
$d_{1} \neq d_{2}$. Consider the two qudit hybrid quantum gate $U_H$ acting on 
$d_{1} \otimes d_{2}$ dimensions, whose action is defined by 
\begin{eqnarray}
U_H \vert m \rangle \otimes \vert n \rangle = \vert m \rangle \otimes \vert m-n \rangle,
\end{eqnarray}
with $m \in \mathbb{Z}_{d_{1}}, n \in \mathbb{Z}_{d_{2}}$ \cite{hybrid}. Let us
take the initial mixed separable state
\begin{eqnarray}
\chi_{d_{1} \otimes d_{2}} = \frac{1}{4}\chi_{x} + \frac{3}{4}\chi_{y},
\end{eqnarray}
 where $\chi_{x}$ is a projector on $\frac{1}{\sqrt{2}}(\vert 0,d_{2}-1 \rangle + \vert 1,d_{2}-1 \rangle)$, and $\chi_{y}$ a projector on $\vert d_{1}-1, d_{2}-1 \rangle$. The state $U_H \chi_{d_{1} \otimes d_{2}} U_H^{\dagger} $ is entangled as identified by the witness $W_{d_{1} \otimes d_{2}}= X^{T_{B}}$ ($ X $ being the projector on $ \vert 02 \rangle - \vert 11 \rangle $). Hence, 
$T_{d_{1} \otimes d_{2}}=U_H^{\dagger} W_{d_{1} \otimes d_{2}} U_H$
 detects $\chi_{d_{1} \otimes d_{2}}$ as a state which is not absolutely separable. 
The above example again illlustrates the fact that one can construct a 
hermitian operator for two qudits (for equal or different dimensions) that can 
recognize useful separable states from which entanglement can be created 
between the two qudits using global unitary operations. 

We finally consider the example of the much discussed {\it CNOT} gate. 
The {\it CNOT} gate can generate entanglement between two qubits, 
if the state under 
consideration is not absolutely separable. If we now consider the action of 
$U_{CNOT}$ on a class of mixed separable states of two qubits of the form 
\begin{equation}
\chi_{mix}=a \vert 00 \rangle \langle 00 \vert + b \vert 00 \rangle \langle 10 \vert + b \vert 10 \rangle \langle 00 \vert + (1-a) \vert 10 \rangle \langle 10 \vert  
\label{mixstate}
\end{equation}
where $a,b \in \mathbb{R}$, we find that the states of the form 
$\chi_{mix}^{e}= U_{CNOT}\chi_{mix} U_{CNOT}^{\dagger}$ can be entangled. Such
entanglement can be detected by the witness 
$W_{CNOT}=
[(\vert 10 \rangle - \vert 01 \rangle)(\langle 10 \vert - \langle 01 \vert)]^{T_{B}}$. 
A hermitian operator $T_{CNOT}$ constructed according to our prescription and which 
detects $\chi_{mix}$ as a state which is not absolutely separable, is given by
$T_{CNOT}=U_{CNOT}^{\dagger}W_{CNOT}U_{CNOT}$.
Now, $Tr(T_{CNOT}\chi_{mix})=-2b$, implying that for $b>0$ the operator detects the 
class of states as useful for entanglement creation under the action of the 
{\it CNOT} gate. For example, if one puts $a=3/4$ and $b=1/4$ for the class
of states (\ref{mixstate}), we get a state that is not absolutely separable
detected by the witness $T_{CNOT}$. On the other hand, a state of the 
form \cite{johnston}
\begin{eqnarray}
\sigma = \frac{1}{11}\left(%
\begin{array}{cccc}
  1 & 0 & 0 & 0 \\
  0 & 3 & 2 & 0  \\
  0 & 2 & 3 & 0  \\
  0 & 0 & 0 & 4  \\
\end{array}%
\right)
\label{mixstate2}
\end{eqnarray}
leads to $Tr(T_{CNOT}\sigma) > 0$, remaining separable under the action of the 
{\it CNOT} gate, as the state $\sigma$ (\ref{mixstate2}) is absolutely 
separable. One may note here though that neither the entanglement witness 
$W_{CNOT}$, and nor consequently $T_{CNOT}$ are universal, as constructed here.
As a result, the operator $T_{CNOT}$ fails to detect some states which are not absolutely separable
that exist even for $b < 0$ in the class of states (\ref{mixstate}). One 
would thus need to construct another 
suitable witness operator to identify states not absolutely separable in the 
latter range.

\paragraph{Decomposition of the witness operator.\textemdash}

For the purpose of experimental determination of the expectation value
of a witness operator on a given state, it is helpful to decompose it in 
terms of spin matrices \cite{decomposition}. 
As an example, the witness $T_{CNOT}$ which detects the class
of states $\chi_{mix}$ (\ref{mixstate}) as not absolutely separable, admits 
the decomposition
$T_{CNOT} = \frac{1}{2}(I \otimes I-I \otimes Z - X \otimes Z - X \otimes I)$
where $X,Z$ are the usual  Pauli spin matrices. Further, in order that the
 witness operator can be measured locally, it may be decomposed in the form
 $T = \sum_{i=1}^{k}c_{i}\vert e_{i}\rangle \langle e_{i}\vert \otimes \vert f_{i}\rangle \langle f_{i}\vert$ \cite{decomposition}. Experimental realization of 
entanglement witnesses has been achieved  using polarized photon 
states \cite{barbieri}. In case of the operator $T_{CNOT}$, the decomposition 
in terms of photon polarization states is given by
\begin{equation}
T_{CNOT}= \vert HV \rangle \langle HV \vert + \vert VV \rangle \langle VV \vert - \vert DH \rangle \langle DH \vert + \vert FH \rangle \langle FH \vert
\end{equation}
where, $\vert H \rangle = \vert 0 \rangle ,\vert V \rangle = \vert 1 \rangle , \vert D \rangle = \frac{\vert H \rangle + \vert V \rangle }{\sqrt{2}}, \vert F \rangle = \frac{\vert H \rangle - \vert V \rangle }{\sqrt{2}}$ are the 
horizontal, vertical and diagonal polarization states 
respectively \cite{barbieri}. The above decomposition suggests a realizable
method to experimentally verify whether it is possible for a mixed separable 
state to give an entangled state on the action of an entangling gate or a 
global unitary operation.

\paragraph{Summary.\textemdash}

In this work we have proposed a framework to distinguish between separable 
states that remain separable from those that become entangled due to global 
unitary operations in any arbitrary dimensional Hilbert 
space \cite{knill,kus,openball,ishizaka,verstraete,hildebrand,johnston}. 
To this 
end we have characterized the set of all absolutely separable bipartite
states as 
convex and compact, enabling one to construct suitable hermitian operators 
for identification of states that are not globally separable. We have suggested
 a generic procedure for construction of such  operators in any dimensions 
which underlines the completeness of the separation, {\it viz.}, if $\chi$ is 
not absolutely separable then there will always be an operator which detects 
it. The action of the operator is demonstrated on states in various
different dimensions. Observational feasibility of witnesses for qubit states
is highlighted 
through decomposition in terms of locally measureable photon polarizations.

The generation of entanglement from separable initial states is of prime
importance in information processing applications \cite{generation}. In 
this context, our
method helps to identify eligible input states for entanglement creation
using global unitary operations in general, and may be of specific relevance 
in quantum gate operations \cite{entangling} widely used in 
quantum computation. Though pure
product states can be readily entangled through such operations, the
inevitability of environmental influences makes the consideration of mixed 
states highly relevant, and thereby lends practical significance to our
proposal for detection of separable mixed states useful for production
of entanglement. Finally, formulations for constructing common and
optimal witnesses analogously to the case of entanglement 
witnesses \cite{lewenstein}, as well as extensions of our scheme for 
multipartite states would be of much relevance.


\begin{thebibliography}{99}

\bibitem{gurvits1} L.Gurvits, \textit{Proceedings of the thirty-fifth annual ACM symposium on Theory of computing}, Eds.  L. L. Larmore and M. X. Goemans, 10 (2003).

\bibitem{peres} A. Peres, Phys. Rev. Lett. \textbf{77}, 1413 (1996).

\bibitem{horodecki} M. Horodecki, P. Horodecki, and R. Horodecki, Phys. Lett. 
A {\bf 223}, 1  (1996).

\bibitem{bound}  P. Horodecki, Phys. Lett. A {\bf 232}, 333 (1997);
M. Horodecki, P. Horodecki, and R. Horodecki, Phys. Rev. Lett. {\bf 80}, 5239 
(1998).

\bibitem{terhal} B. M. Terhal, Phys. Lett. A {\bf 271},  319 (2000).

\bibitem{review} O. Guhne and  G. Toth, Phys. Rep. {\bf 474}, 1  (2009).

\bibitem{barbieri} M. Barbieri, F. De Martini, G. Di Nepi, P. Mataloni, G. M.
 D'Ariano and C. Macchiavello, Phys. Rev. Lett. \textbf{91}, 227901  (2003).

\bibitem{wiec} W. Wieczorek, C. Schmid, N. Kiesel, R. Pohlner, O. Guhne, and 
H. Weinfurter. Phys. Rev. Lett. {\bf 101}, 010503 (2008).

\bibitem{holmes}	R. B. Holmes, {\it Geometric Functional Analysis and its Applications}, (Springer-Verlag,Berlin, 1975).

\bibitem{telwit} N. Ganguly, S. Adhikari, A. S. Majumdar and J. Chatterjee,
Phys. Rev. Lett. \textbf{107}, 270501 (2011).

\bibitem{telwit2} S. Adhikari, N. Ganguly, A. S. Majumdar, Phys. 
Rev. A {\bf 86}, 032315 (2012).

\bibitem{telwitcom} M.-J. Zhao, S.-M. Fei, X. Li-Jost, Phys. Rev. A {\bf 85},
054301 (2012).

\bibitem{knill} E. Knill, \textit{Separability from spectrum}, published electronically at http://qig.itp.uni-hannover.de/qiproblems/15 (2003).

\bibitem{kus} M.Kus and K.Zyczkowski, Phys. Rev. A \textbf{63},032307 (2001).

\bibitem{openball} K.Zyczkowski, P.Horodecki, A. Sanpera and M. Lewenstein, Phys. Rev. A \textbf{58}, 883 (1998): L. Gurvits and H.Barnum, Phys. Rev. 
A \textbf{66}, 062311 (2002).

\bibitem{ishizaka} S. Ishizaka and T. Hiroshima, Phys. Rev. A \textbf{62}, 
022310 (2000).

\bibitem{verstraete} F.Verstraete, K. Audenaert and B.D Moor, Phys. Rev. A \textbf{64}, 012316 (2001).

\bibitem{hildebrand} R. Hildebrand, Phys. Rev. A \textbf{76}, 052325 (2007).

\bibitem{johnston} N. Johnston, Phys. Rev. A \textbf{88}, 062330 (2013).

\bibitem{generation} C.A. Sackett \textit{et al}, Nature \textbf{404}, 256 
(2000); A. Rauschenbeutel \textit{et al}, Science \textbf{288}, 2024 (2000);
 B. Kraus and J. I. Cirac, Phys. Rev. A \textbf{63}, 062309 (2001);
 B. P. Lanyon and N. K. Langford, New J. Phys. \textbf{11}, 013008 (2009);
M.J.Kastoryano \textit{et al} Phy. Rev. Lett. \textbf{106}, 090502 (2011).

\bibitem{thirring} W. Thirring, R. A. Bertlmann, P. Kohler and H. Narnhofer, Eur. Phys. J. D \textbf{64}, 181 (2011).

\bibitem{mixsep} J. Batle \textit{et al}, Phys. Lett. A. \textbf{307}, 253 
(2003); Z. Guan \textit{et al}, arxiv: 1311.5809v1 [quant-ph], to appear in
Phys. Rev. A.

\bibitem{hybrid} J. Daboul, X. Wang and B. C. Sanders, J. Phys. A: Math. Gen. \textbf{36}, 2525 (2003).

\bibitem{entangling} M. J. Bremner \textit{et. al}, Phys. Rev. Lett. 
\textbf{89}, 247902 (2002);  N. Linden, J. A. Smolin and A. Winter, Phys. Rev. 
Lett. \textbf{103}, 030501 (2009); E. T. Campbell, Phys. Rev. A \textbf{82}, 
042314 (2010); M. Musz, M. Kus and K. Zyczkowski, Phys. Rev. A \textbf{87}, 
022111 (2013).

%   ; L. Neves \textit{et al}, Phys. Rev. A \textbf{80},042322(2009).

\bibitem{nmr} S. L. Braunstein \textit{et al}, Phys. Rev. Lett. \textbf{83}, 1054 (1999): D. O. Soares-Pinto, R. Auccaise, J. Maziero, A. Gavini-Viana, R. 
M. Serra, and L. C. Celeri Phil. Trans. Roy. Soc. A {\bf 370}, 4821 (2012).

\bibitem{simmons} G. F. Simmons,\textit{Introduction to Topology and Modern Analysis},(McGraw-Hill, New York, 1963).

\bibitem{isotropic} R. A. Bertlmann, K. Durstberger, B. C. Hiesmayr and P. Krammer, Phys. Rev. A \textbf{72}, 052331 (2005).

\bibitem{decomposition}	O. Guhne, P. Hyllus, D. Bru$\beta$, A. Ekert, M. Lewenstein, 
C. Macchiavello and A. Sanpera, Phys. Rev. A {\bf 66}, 062305 (2002).

\bibitem{lewenstein} M. Lewenstein, B. Kraus, J. I. Cirac, and P. Horodecki, 
Phys. Rev. A {\bf 62}, 052310 (2000); N. Ganguly, S. Adhikari, and A. S. 
Majumdar, Quantum Inf. Process. {\bf 12}, 425 (2013).

\end{thebibliography}
\end{document}